\documentclass[twocolumn,english,prl,showpacs,superscriptaddress]{revtex4}
\usepackage{graphicx,color}
\usepackage{amssymb}
\usepackage[bookmarks]{hyperref}
\usepackage{babel}
\usepackage{esint}

\begin{document}

\title{Electron-plasmon scattering in chiral 1D systems with nonlinear dispersion}

\author{M. Heyl}

\author{S. Kehrein}
\affiliation{Department of Physics, Arnold Sommerfeld Center for Theoretical Physics,
and Center for NanoScience, Ludwig-Maximilians-Universit\"at M\"unchen,
Theresienstr. 37, 80333 Munich, Germany}

\author{F. Marquardt}

\author{C. Neuenhahn}

\affiliation{Department of Physics, Arnold Sommerfeld Center for Theoretical Physics,
and Center for NanoScience, Ludwig-Maximilians-Universit\"at M\"unchen,
Theresienstr. 37, 80333 Munich, Germany}

\affiliation{Institut f\"ur Theoretische Physik, Universit\"at Erlangen-N\"urnberg, Staudtstr. 7, 91058 Erlangen, Germany}

\begin{abstract}
We investigate systems of spinless one-dimensional chiral fermions realized, e.g., in the arms of electronic Mach-Zehnder interferometers, at high energies. Taking into account the curvature of the fermionic spectrum and a finite interaction range, we find a new scattering mechanism where high-energy electrons scatter off plasmons (density excitations). This leads to an exponential decay of the single-particle Green's function even at zero temperature with an energy-dependent rate. As a consequence of this electron-plasmon scattering channel, we observe the coherent excitation of a plasmon wave in the wake of a high-energy electron resulting in the buildup of a monochromatic sinusoidal density pattern. 
\end{abstract}
\maketitle

Many-particle physics in one dimension drastically differs from that
in higher dimensions. In higher dimensions within the scope of Fermi liquid theory, the presence of interactions between fermions does not change the character of the elementary low-energy excitations that are still fermionic. In one dimension this is completely different. Even weak interactions alter the character of the low-energy excitations. They become bosonic and of collective nature. Recently, however, it has been
shown that 1D fermionic systems show Fermi liquid-like behavior at higher energies if
one accounts for the curvature in the spectrum \cite{Khodas}.

\begin{figure}[t]
\includegraphics[width=1\columnwidth]{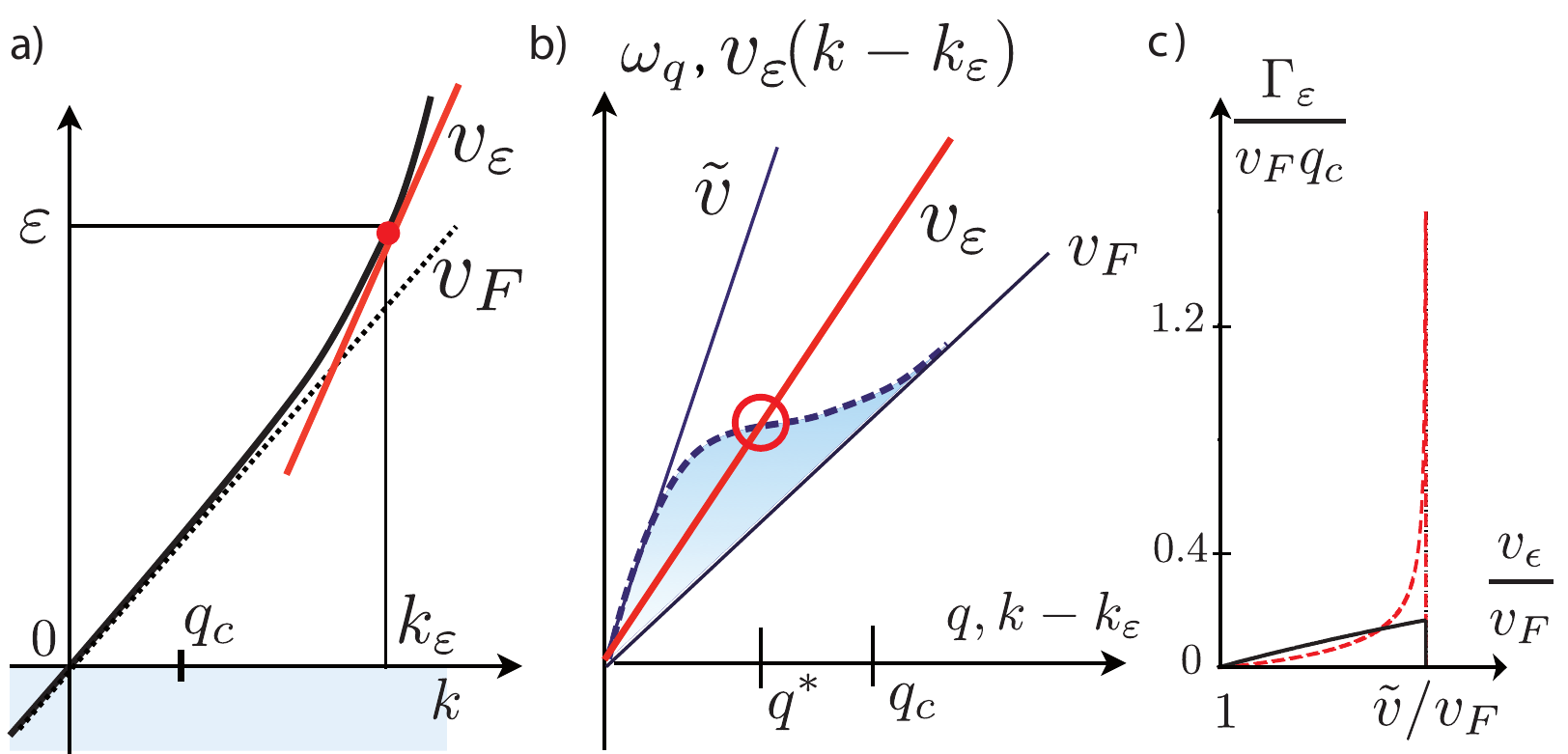}

\caption{(color online) a) High energy fermion injected with energy $\varepsilon\gg q_{c}v_{F}$
on top of the Fermi sea. Due to the curvature of the dispersion it
moves with an energy-dependent velocity $v_{\varepsilon}\geq v_{F}$
b) Sketch of the dispersion relation of the density excitations of
the Luttinger liquid (plasmons) $\omega_{q}$ with the plasmon velocity
$\tilde{v}$ (blue line) and the dispersion of the high-energy fermion
linearized in the vicinity of its initial energy $\varepsilon$ (red
line). The mode $q^{\ast}$ denotes the intersection point of the
two dispersion relations whose existence is responsible for momentum
and energy conserving scattering between the injected electron and
the plasmons. c) Plot of the decay rate of the GF {[}cf. Eq.$\,$(\ref{gammas}){]}
for an analytic interaction potential $U_{q}=2\pi\alpha v_{F}e^{-(q/q_{c})^{2}}$
(dashed line) and a nonanalytic one $U_{q}=2\pi\alpha v_{F}e^{-|q/q_{c}|}$
(solid line), respectively (see main text).\label{fig1} }

\end{figure}

In this work we consider the properties of a system of spinless 1D
chiral fermions under the injection of a high-energy fermion with
well-defined energy $\varepsilon$ beyond the low-energy paradigm.
We take into account the influence of the curvature of the fermionic
dispersion and a finite-range interaction. In experiments, electrons
with well defined energy may be injected via a quantum dot filter
into an integer quantum hall edge state \cite{Feve:2007sf,Altimiras:2009rp}.
Employing these edge channels as the arms of electronic Mach-Zehnder
interferometers \cite{2003_Heiblum_MachZehnder,2006_Neder_MZI,2007_Litvin,2007_Roulleau,2008_Litvin_Lobes},
for example, one may investigate the decoherence of the injected electrons
as a function of injection energy $\varepsilon$. In this regard,
we analyze the Green's function (GF) $G^{>}(x,\varepsilon)=-i\int dt\: e^{i\varepsilon t}\langle\hat{\psi}(x,t)\hat{\psi}^{\dagger}(0,0)\rangle$
(which, in the context of MZI's, is directly related to the interference
contrast \cite{Neuenhahn}), the spectral function $A(k,\varepsilon)$
and the density $\varrho(x,t)$ of the fermionic background in presence
of the high-energy fermion. 

Our main observation is the existence of a new scattering mechanism in chiral 1D systems at high energies due to an interplay of both curvature
and finite interaction range. A fermion injected with a high energy
such that it experiences the curvature of the spectrum scatters
off low-energy density excitations, so-called plasmons. This gives
rise to an exponential decay of the GF in the large distance limit
with a nonzero decay rate $\Gamma_{\varepsilon}$ even at zero temperature 
in stark contrast to the low-energy case where the asymptotic behavior is algebraic.
The excitation of plasmons happens coherently leading to the buildup of a sinusoidal density pattern in the fermionic density in the wake of the injected high-energy electron.

At low energies, interacting 1D fermions are described perfectly well
by a linearized spectrum and a subsequent application of the bosonization
technique. Taking into account curvature one has to employ new methods.
Recently, there has been considerable progress in calculating single-particle
properties beyond the low-energy paradigm \cite{Rozhkov:2005bf,Khodas,Pustilnik:2006kk,Imambekov,Imambekov2,Pereira:2009oj}.
In Ref. \cite{Pustilnik:2006kk} edge singularities in the dynamic
structure factor were found by performing a projection scheme in analogy
to the X-ray edge singularity problem. The authors of~\cite{Imambekov,Imambekov2}
provided a framework for the calculation of response functions for
pointlike interactions beyond the perturbative regime. In a combined
Bethe Ansatz and tDMRG analysis it was shown that the edge behavior
of the spectral function is indeed described by X-ray edge type effective
Hamiltonians and the exact singularity exponents have been determined
\cite{Pereira:2009oj}.

For the calculation of the GF and the spectral function we employ two different methods that turn out to yield exactly the same result. On the one hand, we use a physically transparent semiclassical ansatz whose validity was proven earlier
by comparison to the bosonization result \cite{Neuenhahn}. This ansatz
is naturally extended to include curvature effects. Additionally, we derive an effective Hamiltonian for the description of the single-particle properties by extending the method of Pustilnik \emph{et al.}~\cite{Pustilnik:2006kk} to include the full interaction potential. Based on the latter approach we also obtain the fermionic density after the injection of the high-energy electron.

\emph{Model and GF}. Consider a system of spinless chiral
interacting 1D electrons described by the Hamiltonian\begin{eqnarray}
\hat{H} & = & \sum_{k}\varepsilon_{k}:\hat{c}_{k}^{\dagger}\hat{c}_{k}:+\frac{1}{2}\int dx\, dx'\,\hat{\rho}(x)U(x-x')\hat{\rho}(x')\label{eq:-1}\end{eqnarray}
 where $\hat{\psi}(x)=1/\sqrt{L}\sum_{k}e^{ikx}\hat{c}_{k}$ and we
normal order the Hamiltonian with respect to the vacuum (indicated
by $:\dots:$) where all states with $k<0$ are occupied and empty
otherwise. We denote the fermionic density with $\hat{\rho}(x)=:\hat{\psi}^{\dagger}(x)\hat{\psi}(x):$
and introduce an almost arbitrary interaction potential $U(x)$ with
a Fourier transform $U_{q}\equiv\int dx\, e^{-iqx}U(x)$. The latter
is assumed to be cut-off beyond some momentum scale $q_{c}$, and we
introduce a dimensionless coupling strength $\alpha=U_{q=0}/2\pi v_{F}$.
Whereas the following considerations in principle do not rely on a
particular choice of $\varepsilon_{k}$, for simplicity, we deal with
a dispersion relation of positive curvature as in the case of free
fermions and assume a repulsive interaction, i.e., $\alpha>0$. 

It will be shown below that due to the finite interaction range the indistinguishability between
the injected fermion and the Fermi sea at small temperatures is lifted if the injection energy $\varepsilon\gg v_{F}q_{c}$ is
sufficiently large. This allows for the separation of the high- and
low-energy degrees of freedom, the single fermion propagating ballistically
with the bare velocity $v_{\varepsilon}$ and the remaining fermions
constituting a Tomonaga-Luttinger liquid, respectively. The bosonic
excitations of the latter evolve according to the plasmonic dispersion
relation $\omega_{q}=v_{F}q(1+U_{q}/2\pi v_{F})$ defining the velocity
$\tilde{v}=v_{F}(1+\alpha)$ of the fastest plasmon. 

The fermion and the plasmons are coupled via a residual interaction.
Due to the finite interaction range $1/q_{c}$ and as long as $v_{F}<v_{\varepsilon}<\tilde{v}$
there exists an intersection point $q_{\ast}$ (with $\omega_{q^{\ast}}=v_{\varepsilon}q_{\ast}$)
between the plasmonic spectrum and the dispersion relation of the
single fermion $\varepsilon_{k}-\varepsilon\approx v_{\varepsilon}(k-k_{\varepsilon})$
linearized in the vicinity of its initial energy [see Fig. (\ref{fig1})]. 
The existence of the intersection
point enables an electron-plasmon scattering mechanism conserving
momentum and energy. This manifests in an exponential long-distance
decay of the GF even at zero temperature $T=0$:

\begin{equation}
|G^{>}(x,\varepsilon)|\sim x^{-\gamma_{\varepsilon}}e^{-\Gamma_{\varepsilon}x/v_{\varepsilon}},\: xq_{c}\gg1,\label{eq:exp_decay}\end{equation}
with

\begin{equation}
\Gamma_{\varepsilon}=2\pi^{2}\frac{(v_{\varepsilon}-v_{F})^{2}}{|U_{q_{\ast}}^{'}|}\Theta(\tilde{v}-v_{\varepsilon}),\:\gamma_{\varepsilon}=\left[\frac{\alpha v_{F}}{\tilde{v}-v_{\varepsilon}}\right]^{2}.\label{gammas}\end{equation}
A plot of the decay rate $\Gamma_\varepsilon$ is shown in Fig.~(\ref{fig1}). 
The appearance of the step function $\Theta$ in the expression for
$\Gamma_{\varepsilon}$ mirrors the fact that for $v_{\varepsilon}>\tilde{v}$
the high-energy electron is faster than any plasmonic mode such that
the intersection point $q_{\ast}$ between the plasmonic and fermionic
dispersion relation vanishes [cf. Fig.$\,$(\ref{fig1})]. In
the limit of vanishing curvature, i.e., $v_{\varepsilon}\rightarrow v_{F}$
we have $\Gamma_{\varepsilon}\rightarrow0$ and a power-law exponent
$\gamma_{\varepsilon}\rightarrow1$, which is independent of the coupling
strength $\alpha$ as found earlier \cite{Neuenhahn}. Increasing
the injection energy such that $v_{\varepsilon}\to\tilde{v}$, the
decay rate diverges for analytic interaction potentials. In the limit
of large energies, where $v_{\varepsilon}>\tilde{v}$ and $\Gamma_{\varepsilon}=0$,
the GF decays algebraically for long distances. This decay can be
attributed to the Anderson Orthogonality catastrophe \cite{Anderson:1967gf}
in view of the fact that the GF is the equivalent to the core hole Green's function in the
X-ray edge singularity problem. In this context it is remarkable that the exponent $\gamma_\varepsilon=\Delta n^2$ can be related to the screening charge $\Delta n$, that is the charge displaced
in the fermionic background by the injection of the high-energy fermion.
In the remainder of this article, we will sketch the derivation of
Eq.~(\ref{eq:exp_decay}) and discuss further quantities such
as the spectral function and the density of the fermionic background
after the injection of the high-energy electron. 

\emph{Semiclassical Ansatz for the GF}. Motivated by the earlier results in Ref.~\cite{Neuenhahn} we employ
a semiclassical ansatz for the GF in the limit of large energies $\varepsilon\gg q_{c}v_{F}$.
After its injection, the electron propagates chirally with its bare
velocity $v_{\varepsilon}$, thereby experiencing a fluctuating potential
landscape $\hat{V}(t)=\int dx'\, U(x'-v_{\varepsilon}t)\hat{\rho}_{B}(x',t)$ [see also \cite{Le-Hur:2005fk}]
at its classical position $x=v_{\varepsilon}t$. Here, $\hat{\rho}_{B}(x,t)$
is the fermionic density $\hat{\rho}_{B}(x,t)=L^{-1}\sum_{q>0}\sqrt{n_{q}}(\hat{b}_{q,B}e^{iqx-i\omega_{q}t}+{\rm h.c.})$
[with $(n_{q}=qL/2\pi)$] of the bath electrons with bosonic operators
$\hat{b}_{q,B}=1/\sqrt{n_{q}}\sum_{k}\hat{c}_{k,B}^{\dagger}\hat{c}_{k+q,B}$
representing the plasmonic excitations evolving according to the plasmonic
dispersion $\omega_{q}$.
It is assumed that the non-linearity of
the fermionic dispersion is small enough such that the velocity of
the propagating fermion can be considered as constant and the remaining
electrons can be treated by means of bosonization. 
Specifically, the change in velocity of an electron due to a scattering event with a typical momentum transfer $q_c$ has to be small such that $q_c\partial^2\varepsilon_k/\partial k^2\ll \partial \varepsilon_k/\partial k$ for all momenta $k$ near the Fermi momentum and near $k_\varepsilon$.
As a consequence
of the fluctuating plasmonic quantum bath, the high-energy fermion
accumulates a random phase and its non-interacting GF is multiplied
by the average value of the corresponding phase factor:

\begin{eqnarray}
\frac{G^{>}(x,\varepsilon)}{G_{0}^{>}(x,\varepsilon)} & = & \left\langle \hat{T}\exp\left[-i\int_{0}^{x/v_{\varepsilon}}dt'\,\hat{V}(t')\right]\right\rangle .\label{eq:-3}\end{eqnarray}
Here, $\hat{T}$ denotes the time ordering symbol and $G_{0}^{>}=-ie^{ik_{\epsilon}x}/ v_{\varepsilon}$
is the non-interacting GF for $\varepsilon>0$. Note that the whole
influence of the finite curvature is contained in the energy dependence
of $v_{\varepsilon}\geq v_{F}$. Employing the Gaussian nature of
the plasmonic bath it is possible to express the r.h.s. of Eq.~(\ref{eq:-3})
in terms of the auto-correlation function of the potential fluctuations
$\langle\hat{V}\hat{V}\rangle_{\omega}=\int dt\, e^{i\omega t}\langle\hat{V}(t)\hat{V}(0)\rangle$
(cf.$\,$\cite{Neuenhahn}) experienced by the single electron in
its co-moving frame of reference. In particular, for the modulus of
the GF one obtains
\begin{eqnarray}
\left|\frac{G^{>}}{G_{0}^{>}}\right| & = & \exp\left[-\int_{-\infty}^{\infty}\frac{d\omega}{2\pi}\,\frac{\sin^{2}(\omega x/2v_{\epsilon})}{\omega^{2}}\langle\{\hat{V},\hat{V}\}\rangle_{\omega}\right],\label{eq:-4}\end{eqnarray}
where only the symmetrized correlator $\langle\{\hat{V},\hat{V}\}\rangle_{\omega}=\langle\hat{V}\hat{V}\rangle_{\omega}+\langle\hat{V}\hat{V}\rangle_{-\omega}$
enters. The asymptotic long-distance decay of the GF in Eq.~(\ref{eq:-2})
is governed by the low-frequency properties of the potential fluctuation
spectrum $\langle\{\hat{V},\hat{V}\}\rangle_{\omega\downarrow0}=2\pi\gamma_{\varepsilon}|\omega|+4\Gamma_{\varepsilon}$
(here we took $T=0$). It consists of an Ohmic part responsible for
the power-law decay and a constant offset which leads to an exponential
decay of the GF. For intermediate energies where $v_{F}q_{c}\ll\varepsilon$
but $v_{\varepsilon}=v_{F}$, Eq.~(\ref{eq:-3}) reproduces exactly
the GF from standard bosonization~\cite{Neuenhahn}. Thus, our analysis is correct within the validity of the bosonization technique.

\emph{Effective Hamiltonian}.  In fact, the semiclassical ansatz for
the GF in Eq.~(\ref{eq:-3}) matches precisely the result obtained
by  an extension of the approach  in Ref.~\cite{Pustilnik:2006kk}
to treat the full interaction potential. There, $G^{>}(x,t)$ is viewed
as an impurity problem related to the X-ray edge singularity \cite{Nozieres}
where a scatterer, the injected fermion in this case, is suddenly
switched on. The Hamiltonian in Eq.~(\ref{eq:-1}) is projected onto
two strips of states that capture the relevant degrees of freedom,
an energy interval around the initial energy $\varepsilon$ of the
injected high-energy fermion labeled by $\varepsilon$ and an energy
window around the Fermi energy labeled by an index $B$. Under this
projection, the fermionic field decomposes into $\hat{\psi}(x)\to\hat{\psi}_{B}(x)+e^{ik_{\varepsilon}x}\hat{\psi}_{\varepsilon}(x)$.
Linearizing the dispersion relations within both strips of states,
the Hamiltonian of the low-energy sector can be bosonized. Regarding
correlation functions involving at most one high-energy electron one
obtains: \begin{eqnarray}
H=\sum_{q>0}\omega_{q}b_{q,B}^{\dag}b_{q,B}+\int dx\,\hat{\psi}_{\varepsilon}^{\dag}(x)(\varepsilon-iv_{\varepsilon}\partial_{x})\hat{\psi}_{\varepsilon}(x)\nonumber \\
+\int dxdx'\hat{\rho}_{B}(x)U(x-x')\hat{\psi}_{\varepsilon}^{\dagger}(x')\hat{\psi}_{\varepsilon}(x').\label{eq_eff_Hamiltonian}\end{eqnarray}
 In the derivation of this effective Hamiltonian a contribution
proportional to $U_{k_{\varepsilon}}$ has been neglected as $k_{\varepsilon}\gg q_{c}$
and $U_{q}$ rapidly decays for $q\gg q_{c}$ by assumption. The omitted
term is responsible for exchange processes between the high- and low-energy
sector lifting the distinguishability between the high-energy fermion
and the low-energy degrees of freedom. In Eq.~(\ref{eq_eff_Hamiltonian}) 
a constant Fock shift $U(x=0)\hat{N}_{\varepsilon}/2$ is omitted
which drops out automatically if we take as a starting point the Coulomb
interaction instead of the density-density interaction in Eq.~(\ref{eq:-1}).
The Hamiltonian in Eq.~(\ref{eq_eff_Hamiltonian}) can be diagonalized by
means of the unitary transformation $\hat{U}=\exp(\hat{S})$ where
$\hat{S}=\int dx \hat{\psi}_{\varepsilon}(x)\hat{\psi}_{\varepsilon}(x)\sum_{q>0}\chi_{q}[\hat{b}_{q,B}^{\dagger}e^{-iqx}-H.c.]$
with $\chi_{q}=2\pi U_{q}/(U_{q}-2\pi(v_{\varepsilon}-v_{F}))\sqrt{n_{q}}$.
From Eq.~(\ref{eq_eff_Hamiltonian}), one can calculate the golden
rule rate for the excitation of plasmons by the high-energy fermion.
It matches precisely the decay rate
determining the exponential decay of the GF in Eq.~(\ref{gammas}). 

\emph{Spectral function}.
\begin{figure}
\centering \includegraphics[width=0.9\columnwidth]{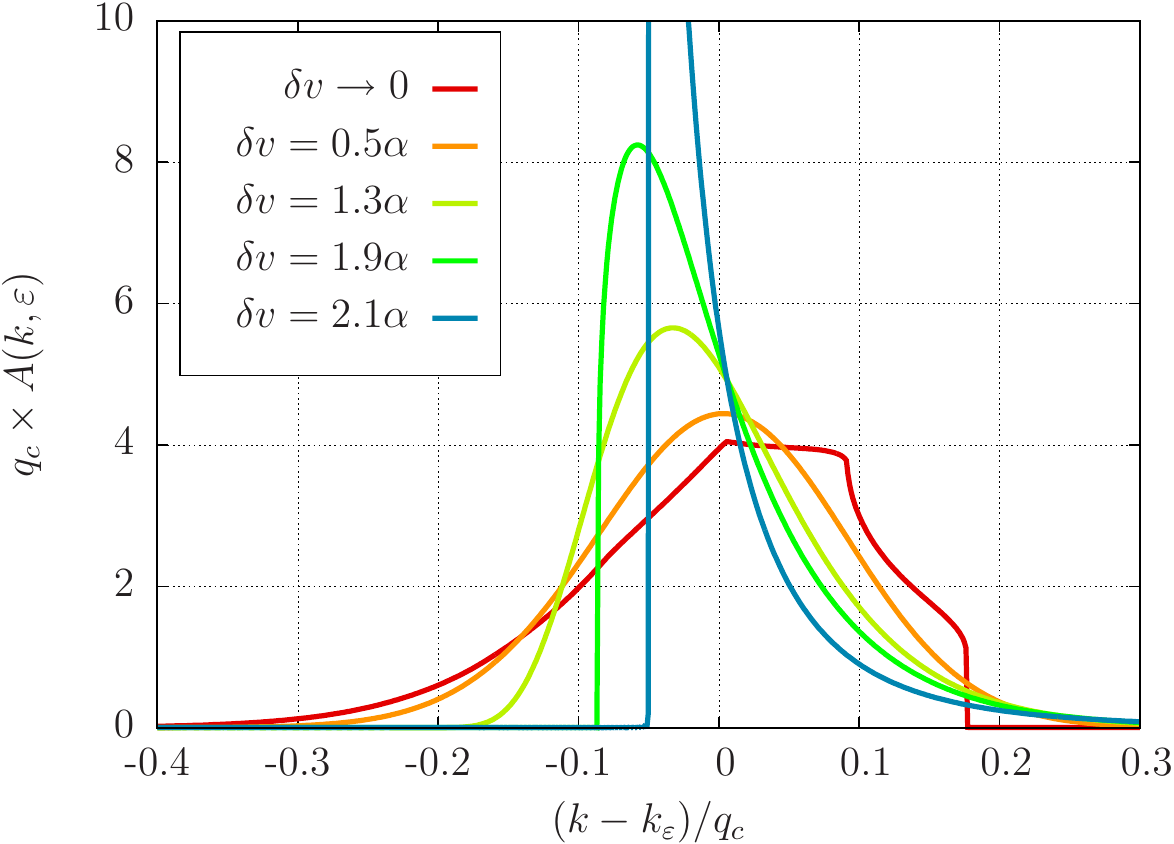} 

\caption{(color online) Spectral function $A(k,\varepsilon)$ for different velocities $v_{\varepsilon}$
of the high-energy fermion where $\delta v=(v_\varepsilon-v_F)/v_F$. For these plots we have chosen an analytic potential $U_{q}=2\pi v_{F}\alpha\exp[-(q/q_{c})^{2}]$
with $\alpha=0.2$.}

\label{pic_A} 
\end{figure}
The spectral function $A(k,\varepsilon)$ is connected to the GF via
$A(k,\varepsilon)=i/(2\pi)\int dx\, e^{-ikx}G^{>}(x,\varepsilon)$
($\epsilon>0$ and $T=0$). It behaves qualitatively different whether
the exponent $\gamma_{\varepsilon}$ appearing in the large distance
behavior of the GF is bigger or smaller than one. Remarkably, this
property is not connected to the distinction between $v_{\varepsilon}\lessgtr\tilde{v}$
that determines the threshold between exponential and algebraic large
distance behavior for the GF. For $\gamma_{\varepsilon}<1$ or equivalently
$\delta v=(v_{\varepsilon}-v_F )/v_F>2\alpha$, the spectral function shows a
power-law singularity together with a threshold behavior for $k\to\kappa_{\varepsilon}+k_{\varepsilon}$
\begin{eqnarray}
A(k,\varepsilon) & \sim & \sin(\gamma_{\varepsilon}\pi/2)\left(k-\kappa_{\varepsilon}-k_{\varepsilon}\right)^{\gamma_{\varepsilon}-1}\theta\left(k-\kappa_{\varepsilon}-k_{\varepsilon}\right)\label{eq:-2}\end{eqnarray}
where $\kappa_{\varepsilon}=v_{\varepsilon}^{-1}\mu_{\varepsilon}$
and $\mu_{\varepsilon}=\int_{0}^{\infty}dq(U_{q}/2\pi)^{2}/(v_{\varepsilon}-v_{F}-U_{q}/2\pi)$
denotes the energy that is needed to overcome the Coulomb interaction while injecting an electron with
energy $\varepsilon$. In Fig.~(\ref{pic_A}), the curve with $\delta v=2.1 \alpha$ shows the spectral function with a power law singularity according to Eq.~(\ref{eq:-2}). Note that the support of the spectral function,
$A(k,\varepsilon)\not=0$ only for $\varepsilon+\mu_\varepsilon<\varepsilon_{k}$,
is exactly opposite to the low-energy Tomonaga-Luttinger Liquid case
where $A(\varepsilon,k)\not=0$ only for $\varepsilon+\mu_\varepsilon>\varepsilon_{k}$.
This is a consequence of the condition $v_{\varepsilon}>\tilde{v}$
implying that an electron with wave vector $k$ can excite plasmonic
modes only by reducing the energy in the system [see Fig.~\ref{fig1}].
In the limit $\varepsilon\to\infty$ where $\gamma_{\varepsilon}\to0$,
one recovers the free particle, a $\delta$-function in the spectrum
as $\lim_{\eta\to0}\eta|x|^{\eta-1}/2=\delta(x)$. As shown in Ref.
\cite{Schonhammer} this is not the case for a linearized dispersion
even in the limit $\varepsilon\to\infty$.

For $\gamma_{\varepsilon}>1$, i.e. $v_{F}<v_{\varepsilon}<2\tilde{v}-v_{F}$,
the spectrum changes drastically. The singularity vanishes and the
spectral function merely becomes a skew Gaussian, compare Fig.~(\ref{pic_A}). In
the regime $\gamma_{\varepsilon}>1$ the GF, that is the Fourier transform of $A(k,\varepsilon)$, is dominated by its initial
Gaussian decay due to strong dephasing by the plasmonic background fluctuations. Thus, the spectrum itself is also dominated by the incoherent background such that no well-defined quasiparticle peak is visible in spite of the exponential decay of the GF.
In the limit $v_{\varepsilon}\to v_{F}$
and for a potential $U_{q}$ with a sharp cutoff at $q_{c}$, one
recovers the result by Ref.~\cite{Schonhammer} as indicated in Fig.~(\ref{pic_A}).

\emph{Coherent emission of plasmon waves}. 
\begin{figure}
\includegraphics[width=1\columnwidth]{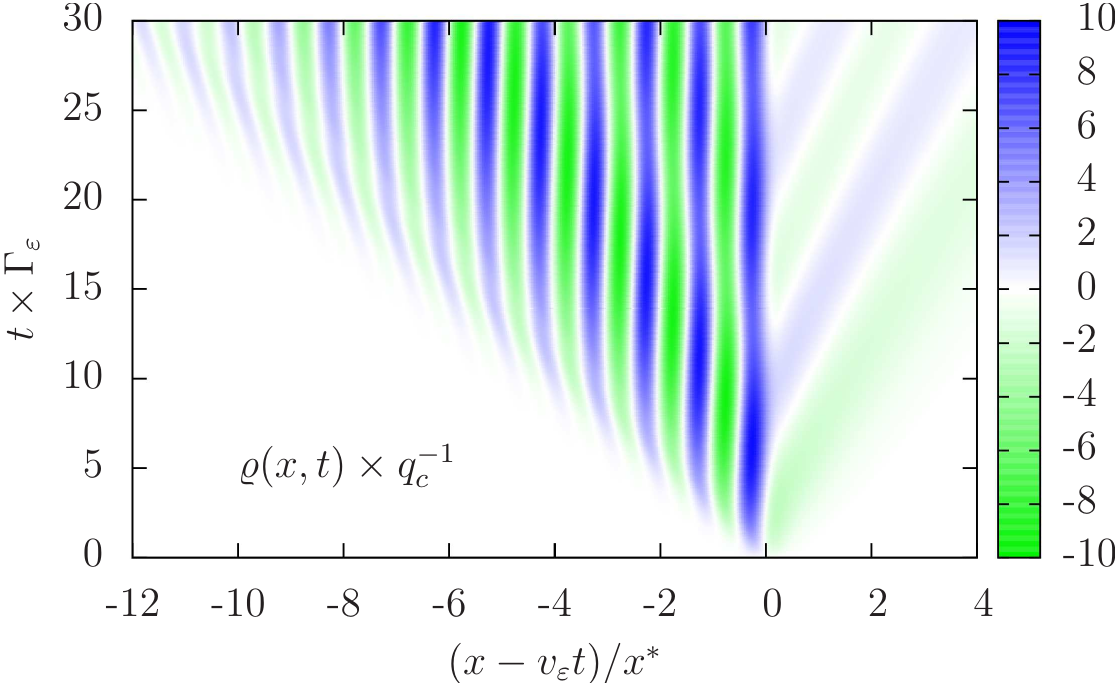}

\caption{(color online) The fermionic density $\varrho(x,t)$ in the co-moving frame of the
high-energy electron injected at time $t=0$. The period $x^{*}$ of the oscillations is given
by $x^{\ast}=2\pi/q_{\ast}$. For this plot we have chosen the same
potential $U_{q}$ as in Fig. \ref{pic_A} and $\delta v=0.5 \alpha$.\label{fig:CDM}}
\end{figure}
In order to investigate the influence of the electron-plasmon scattering mechanism onto the fermionic background, we analyze the fermionic density of the bath \begin{equation}
\varrho(x,t)=\mathcal{N}\langle\psi_{0}|\hat{\psi}_{\varepsilon}(0)\hat{\rho}_B(x,t)\hat{\psi}_{\varepsilon}^{\dagger}(0)|\psi_{0}\rangle\label{eq:}\end{equation}
in the presence of the high-energy electron. Here, $|\psi_{0}\rangle$
is the ground state of the Hamiltonian in Eq.~(\ref{eq_eff_Hamiltonian})
without the high-energy fermion and $\mathcal{N}=\langle \psi_0 | \hat{\psi}_\varepsilon(0) \hat{\psi}_\varepsilon^\dagger(0) | \psi_0\rangle^{-1}$ a normalization
constant. In the parameter regime $v_{\varepsilon}<\tilde{v}$ where
electron-plasmon scattering takes place one observes the coherent
emission of plasmon waves with wave vector $q_{\ast}$  of the resonant plasmonic mode [see Fig.$\,$(\ref{fig:CDM})].
In the limit $t\to\infty$ and for distances $x-v_{\varepsilon}t\gg q_{c}^{-1}$
sufficiently far away from the position of the high-energy electron,
we obtain the following analytic result:

\begin{equation}
\varrho(x,t)\rightarrow\Theta(v_{\varepsilon}t-x)\sin[q_{\ast}(x-v_{\varepsilon}t)]U_{q_{*}}/U_{q_{\ast}}^{'}\label{eq:-5}\end{equation}
As can be seen in Fig.~(\ref{fig:CDM}), the coherent density
excitations build up within a 'light cone' $x\in[v_{F}t,\tilde{v}t]$
\cite{Calabrese:2006sf} set by the minimal and maximal plasmonic
phase velocities. The wavelength of the oscillations
in the density $\varrho(x,t)$ is tunable by the choice of an appropriate
injection energy $\varepsilon$ of the high-energy fermion. 

For velocities $v_{\varepsilon}>\tilde{v}$, no scattering between
electrons and plasmons is possible. In this case, the density $\varrho(x,t)$
can be separated into two contributions. The first one describes the
initial excitation of plasmonic modes right after the injection of
the high-energy fermion. This transient perturbation cannot follow
the electron that is faster than any plasmonic mode. The second contribution
traveling together with the high-energy electron is responsible for
the screening of the injected charge and is reminiscent of viewing
the GF as an impurity problem. Integrating over space then provides
us with the screening charge $\Delta n$, the charge displaced by
the introduction of the local scatterer. As mentioned before, it is directly related to the exponent $\gamma_{\varepsilon}=\Delta n^{2}$ of the GF.

\emph{Conclusions}.  We have discussed electron-plasmon scattering
in systems of 1D chiral electrons. This scattering leads to an exponential
decay of the single-particle Green's function even at zero temperature
and to a coherent monochromatic pattern in the fermionic density in
the wake of the electron. This effect is absent in the low-energy
limit and relies exclusively on the interplay between a finite interaction
range and a non-linear fermionic dispersion. 

\emph{Acknowledgements}.  We thank L. Glazman for fruitful discussions.
Financial support by NIM, CeNS, the Emmy-Noether program and the SFB/TR
12 is gratefully acknowledged.

\bibliographystyle{apsrev}
\bibliography{literature}

\end{document}